\documentclass{aastex63}

\usepackage{epsfig}          
\usepackage{graphicx}        
\usepackage{natbib}         
\usepackage{amssymb}        
\usepackage[T1]{fontenc}        

\newcommand{\SMM}{{\em SMM}\/}

\newcommand{\caxix}{{Ca~{\sc xix}}}
\newcommand{\caxviii}{{Ca~{\sc xviii}}}

\newcommand{\fexxi}{{Fe~{\sc xxi}}}

\newcommand{\fexxv}{{Fe~{\sc xxv}}}
\newcommand{\fexxvi}{{Fe~{\sc xxvi}}}

\received{...}
\revised{...}
\accepted{...}
\submitjournal{ApJ}

\shorttitle{SMM Bent Crystal Spectrometer Collimator}
\shortauthors{J. Sylwester et al.}

\begin{document}



\title{A Unique Resource for Solar Flare Diagnostic Studies: the {\em SMM}\/ Bent Crystal Spectrometer}

%
\correspondingauthor{K. J. H. Phillips}
\email{kennethjhphillips@yahoo.com}


\author[0000-0002-8060-0043]{J. Sylwester}
\affiliation{Space Research Centre, Polish Academy of Sciences (CBK PAN), Warsaw, Bartycka 18A, Poland}

\author[0000-0001-8428-4626]{B. Sylwester}
\affiliation{Space Research Centre, Polish Academy of Sciences (CBK PAN), Warsaw, Bartycka 18A, Poland}

\author[0000-0002-3790-990X]{K. J. H. Phillips}
\affiliation{Scientific Associate, Earth Sciences Dept., Natural History Museum, Cromwell Road, London SW7 5BD, UK}

\author[0000-0002-5299-5404]{A. K\k{e}pa}
\affiliation{Space Research Centre, Polish Academy of Sciences (CBK PAN), Warsaw, Bartycka 18A, Poland}

\author[0000-0002-2858-3661]{C. G. Rapley}
\affiliation{Dept. of Earth Sciences, University College London, Gower Street, London WC1E 6BT, UK}


\begin{abstract}
The {\em Bent Crystal Spectrometer}\/ (BCS) on the NASA {\em Solar Maximum Mission}\/ spacecraft observed the X-ray spectra of numerous solar flares during the periods 1980 February to November and 1984~--~1989. The instrument, the first of its kind to use curved crystal technology, observed the resonance lines of He-like Ca (\caxix) and Fe (\fexxv) and neighboring satellite lines, allowing the study of the rapid evolution of flare plasma temperature, turbulence, mass motions etc. To date there has not been a solar X-ray spectrometer with comparable spectral and time resolution, while subsequent solar cycles have delivered far fewer and less intense flares. The BCS data archive thus offers an unparalleled resource for flare studies. A recent re-assessment of the BCS calibration and its operations is extended here by using data during a spacecraft scan in the course of a flare on 1980 November~6 that highlights small deformations in the crystal curvature of the important channel~1 (viewing lines of \caxix\ and satellites). The results explain long-standing anomalies in spectral line ratios which have been widely discussed in the past. We also provide an in-flight estimation of the BCS collimator field of view which improves the absolute intensity calibration of the BCS. The BCS channel~1 background is shown to be entirely due to solar continuum radiation, confirming earlier analyses implying a time-variable flare abundance of Ca. We suggest that BCS high-resolution \caxix\ and \fexxv\ line spectra be used as templates for the analysis of X-ray spectra of non-solar sources.
\end{abstract}

%
\keywords{atomic data -- Sun: abundances --- Sun: corona --- Sun: flares --- Sun: X-rays, gamma rays}

%
\section{Introduction}\label{sec:Intro}

The NASA {\em Solar Maximum Mission}\/ (\SMM) spacecraft, launched on 1980 February~14 at the peak of solar cycle~21, carried a battery of instruments to observe the active, particularly the flaring, Sun at ultraviolet, X-ray, and gamma-ray wavelengths. It remained operational until 1989, although there was a loss of fine pointing between 1980 November and 1984 April due to the failure of an attitude control unit which was replaced by astronauts on the {\em Space Shuttle SMM Repair Mission}  (STS-41-C).

The {\em Bent Crystal Spectrometer}\/ on \SMM\ was a high-resolution spectrometer designed to observe the X-ray spectra of solar flares in the region of highly ionized Ca and Fe emission lines. The intensity ratios of the lines determine a range of physical parameters for the hot flare plasma including temperature, emission measure, and plasma motions. The instrument  formed a key component of the X-ray Polychromator \citep{act80}. It had eight channels each consisting of a bent germanium crystal wafer and a position-sensitive proportional counter, enabling spectra in narrow ranges to be instantaneously formed by Bragg diffraction over time intervals short enough to follow the changing flare intensity, particularly during the impulsive stage. The BCS pioneered the concept of curved crystal technology for space instrumentation, the advantage of which was the lack of a need for moving mechanical parts as with scanning flat crystal spectrometers as well as encoders to read the scanning angle. The BCS was fixed on the spacecraft baseplate, so that particular active regions could be selected through spacecraft pointing, which was done on a day-by-day preplanned basis. A grid collimator confined the emission from a single active region to be viewed, avoiding spectral confusion when two or more flares occurred simultaneously in different regions.

An extensive review of the BCS instrumental parameters was made by \cite{rap17} so that more accurate analysis of the data can now be made. There is continued interest in the BCS data as the spectral resolution (resolving power $\lambda / \Delta \lambda \sim 4000$ for its long-wavelength channel viewing \caxix\ lines but up to $15000$ for three of its high-resolution \fexxv\ channels) for such a long-operating instrument was unprecedented and continues to be unsurpassed for any solar X-ray spectrometer. To our knowledge, only the SOX2 X-ray spectrometer on {\em Hinotori}\/ \citep{tan86}, which viewed \fexxv\ and \fexxvi\ spectra during flares over a short period in 1981~--~1982, had better spectral resolution. The BCS was used to observe the upflowing plasma and turbulence occurring at the flare impulsive stage, leading to flare models invoking chromospheric evaporation processes \citep{ant82,cul81}. The replacement of the \SMM\ attitude control unit and other repairs during the 1984 Space Shuttle repair mission enabled the front panels of the various {\em SMM}\/ instruments to be inspected, and it was evident, from photographs taken by the astronauts, that the thermal filter of the BCS had disintegrated at some point since launch in 1980. Careful analysis of data from various flares by \cite{rap17} now indicates that this failure occurred on or around 1980 October~14, and is probably associated with the large ({\em GOES}\/ class X3) flare that occurred on this date. Calibration corrections have been calculated and are applied to the spectra to account for this change.

In this work, we extend the analysis of \cite{rap17} by considering BCS data collected during a unique data set made over the daylight period of an orbit (approximately one hour) on 1980 November~6 when the \SMM\ spacecraft executed a series of maneuvers in which a flaring active region was repeatedly scanned in a square pattern, approximately six arcminutes on a side. The flare emission thus passed over the BCS collimator both aligned with and perpendicular to the BCS dispersion direction. In each case, the motions modulated the intensity of the spectra observed according to the angular response function of the collimator. The motions aligned with the dispersion axis gave rise to spectra that appeared to shift along the BCS detectors as a result of the changing angle of incidence (Section~\ref{sec:BCS_Sp_SMM_scans}). The analysis reported here is particularly for BCS channel~1, which viewed a group of lines due to He-like Ca (\caxix) and associated satellite lines, because these spectra have the highest count rates of all eight channels and the \caxix\ line emission persisted for much longer than the \fexxv\ emission. From channel~1 spectra we found evidence for irregularities in the crystal curvature that then explained a long-standing problem in the intensity ratio of the \caxix\ $x$ and $y$ intercombination lines (see Table~\ref{tab:line_features} and Figure~\ref{fig:three_BCS_sp} for identification of the spectral lines of the BCS channel~1 spectra). The angular extent of the BCS collimator was also estimated, confirming and improving on pre-flare measurements. Fits to BCS channel~1 spectra using theoretical spectra, similar to those used in recent analysis of {\em CORONAS-F}\/ DIOGENESS spectra \citep{phi18}, depend on electron temperature, mainly through the ratio of the \caxviii\ satellite line $k$ to the \caxix\ resonance line $w$, and it was found that this temperature is similar to that estimated from the two channels of {\em GOES}\/, so offering a useful proxy for temperatures using \caxix\ line ratios. This in turn helped us to derive a new absolute intensity calibration for channel~1 from the comparison of BCS emission with the fluxes of the two channels of {\em GOES}\/.  We also deduced that the channel~1 background pedestal can be attributed to solar continuum with negligible amounts of instrumental (e.g. crystal fluorescence) background, which confirms an assumption made in previous studies of the time-varying flare abundance of calcium \citep{jsyl84,jsyl98}.

\section{The Bent Crystal Spectrometer}
\label{sec:BCS}

\begin{figure}
\centerline{\includegraphics[width=0.6\textwidth,clip=]{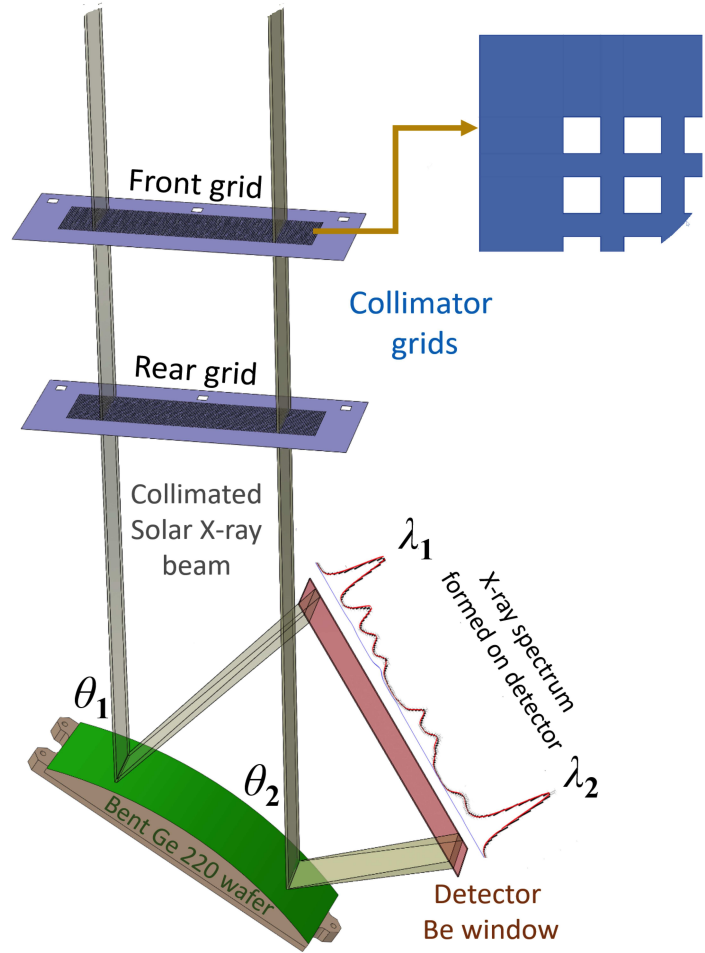}}
\caption{Ray path scheme for BCS channel~1 (\caxix) generated by Computer Aided Design (CAD). Solar X-rays (parallel gray bars, upper left) are incident through a thermal filter and the $6 \times 6$~arcmin (FWHM) square multi-grid collimator (front and rear grids indicated) on to the Ge~220 cylindrically bent crystal wafer. The diffracted rays (pale gray bars), incident on the bent crystal at angles $\theta_1$ and $\theta_2$, have different wavelengths ($\lambda_1$, $\lambda_2$) according to their position along the crystal. The dispersed radiation received by the position-sensitive detector forms a complete spectrum over the wavelength range 3.169~--~3.227~\AA\ for a flare on the BCS optical axis defined by the direction of the collimator's peak transmission.
\label{fig:BCS_ray_scheme}}
\end{figure}

An outline of the principle of the BCS is shown in Figure~\ref{fig:BCS_ray_scheme}. Incoming X-rays from a solar flare were incident through the BCS collimator and thermal filter on to one of eight curved crystal wafers mounted on substrates. The rays were then diffracted according to Bragg's law, $\lambda = 2d\,\, {\rm sin}\,\, \theta$ ($\lambda = $ diffracted ray wavelength, $\theta = $ grazing angle of incidence, $d = $ crystal lattice spacing). Because of the crystal curvature, slightly different values of $\theta$ occur for different points along each crystal's length. Eight position-sensitive proportional counters received the diffracted radiation from each crystal, so over short data gathering intervals, typically a few seconds, complete spectra were formed as the detected photons were read off from the anode wire of each detector. Figure~\ref{fig:BCS_ray_scheme} illustrates the scheme for BCS channel~1, the wavelength range of which includes the \caxix\ resonance line and associated satellites and for which the diffracting crystal was a thin wafer of Ge~220 cylindrically bent crystal with $2d = 4.000$~\AA\ and nominal radius of curvature equal to 5.83~m. A source like a solar flare or hot active region that was off-axis by an angle $\Delta \theta$ along the crystal dispersion direction, or solar E~--~W during nearly all spacecraft operations, would give rise to a spectrum shifted in wavelength by an amount $\Delta \lambda$ given by

\begin{equation}
\Delta \lambda = 2d\, {\rm cos}\, \theta \,\Delta \theta.
\label{eq:diff_Bragg}
\end{equation}

The BCS collimator, operating on the standard Oda principle \citep{oda65}, consisted of eight arrays of co-aligned square holes ($0.350 \times 0.350$~mm) evenly located to form a $\sim 6 \times 6$~arcmin (FWHM) square pyramid transmission pattern (Section~\ref{sec:BCS_Coll_Ang_Res}). Only the front and rear grids are shown in Figure~\ref{fig:BCS_ray_scheme}. Based on its design and on pre-flight optical and X-ray calibrations, the estimated on-axis transmission is $33\pm 3$\,\%. Offsets of the flare source from the instrument optical axis defined by the direction of the peak transmission of the collimator accordingly result in intensity modulations.

%

\section{1980 November~6 Flare Images}
\label{sec:flare_obs}

\begin{figure}
\centerline{\includegraphics[width=0.7\textwidth,clip=]{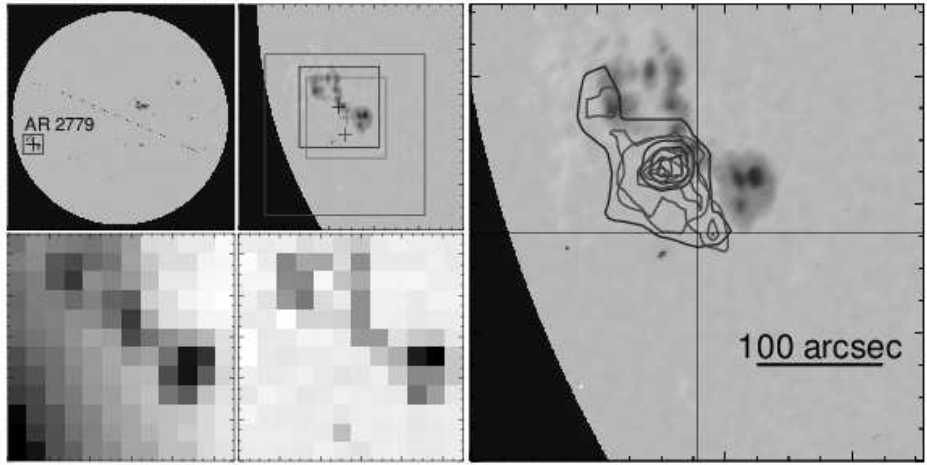}}
\caption{Top left: White-light image of the Sun  from Kislovodsk Observatory taken on 1980 November~7. Solar North is up, east to the left. Top middle: Sunspots of AR~2779 with fields of view of \SMM\ instruments: the red square box shows the BCS collimator field of view (red cross is the BCS optical axis), the blue square the FCS scan dimensions, and the green square the UVSP scan dimensions. Bottom left: FCS white-light sensor image of the sunspots of AR~2779. Bottom middle: Kislovodsk sunspot image spatially degraded to match the FCS image. Right-hand panel: The Kislovodsk sunspot image with overlaid FCS flare images (blue: O~{\sc viii} contours; red: S~{\sc xv}) and UVSP \fexxi\ flare image (green contours). The position of the BCS optical axis is at the intersection of the continuous lines.
\label{fig:sunspot_images}}
\end{figure}

BCS data from the M3.5 flare on 1980 November~6 (SOL1980-11-06T22:27) are of particular interest since the \SMM\ spacecraft performed raster-like scans over active region AR~2779 during the flare enabling BCS line shifts to be precisely related to the spatial shifts. We estimated the solar location of the flare and the host active region using imaging data from two \SMM\ instruments: the {\em Flat Crystal Spectrometer}\/ (FCS) observing soft X-rays and white-light emission using an in-built sensor; and the {\em Ultraviolet Spectrometer and Polarimeter}\/ (UVSP), which was an imaging spectrometer observing in the 1750~--~3600~\AA\ range. White-light images from the Kislovodsk Observatory (part of the Pulkovo Observatory, St Petersburg, Russia), using the Debrecen catalogue (see \cite{gyo17})\footnote{Web site is www.fenyi.solarobs.csfk.mta.hu/DPD} are given in Figure~\ref{fig:sunspot_images}, taken at around 06:00~UT on November~7, some eight hours after the M3.5 flare peak. A full-Sun image and a more detailed image around the sunspots of AR~2779 are shown (top left panels and right panel). The FCS white-light sensor images (spatial resolution $\sim 14$~arcseconds) were correlated with the Kislovodsk sunspot image by degrading the spatial resolution of the latter (lower panels). The flare maximum X-ray and ultraviolet emission are shown relative to the Kislovodsk sunspot image in the right panel. The FCS showed significant emission during the flare rise from five of its seven X-ray channels, the lowest-temperature ($\sim 3.5$~MK) emission being from H-like oxygen (O~{\sc viii}, 18.97~\AA, blue contour, right panel), and the highest-temperature emission from He-like sulfur (S~{\sc xv}, temperature $\sim 16$~MK, red contour). The UVSP obtained images in the region of 1300~\AA; the wavelength range included the forbidden line of C-like Fe (\fexxi, 1354~\AA, green contours: see \cite{dos75}), with similar emitting temperature to the S~{\sc xv} X-ray line. The \fexxi\ and S~{\sc xv} emission regions are, as expected from their similar temperature dependence, nearly identical in extent.

%

\section{\SMM\ Spacecraft maneuvers}
\label{SMM_maneuvers}

Throughout the \SMM\ spacecraft maneuvers during the 1980 November~6 flare, BCS spectra were obtained with data gathering interval equal to 19~s. In a pre-planned sequence, the hour-long daylight portion of the \SMM\ orbit started with the spacecraft pointed at AR~2779 for a period of 13 minutes starting at 22:02~UT. The FCS performed twenty-three rasters, $3 \times 3$~arcmin square, over AR~2779. From {\em GOES}\/ light curves, the M3.5 flare began its rise over this period, referred to here as Phase~$A$. At 22:15~UT, the spacecraft started its series of raster maneuvers. The {\em GOES}\/ light curves indicate that over this period (Phase~$B$) the flare reached its maximum at about 22:26~UT, with a further smaller peak at 22:32~UT. The spacecraft raster motion had angular excursions of $\pm 210$~arcseconds in the E~--~W  direction ($\pm x$ direction according to the usual solar coordinate convention, with $+x$ being the W direction), i.e. the spectrometer's dispersion axis, and with each excursion taking 240~seconds. As explained in Section~\ref{sec:BCS_Coll_Ang_Res}, the spacecraft also moved northwards at the end of each E~--~W scan, forming a ``boustrophedonic'' (``as the ox ploughs'') pattern.

Although official documentation on the nature of the spacecraft maneuvers is no longer available, it is known that they occurred through the activation of on-board reaction wheels in the \SMM\ attitude control unit. (A failure in this unit later in 1980 November was the reason that the \SMM\ pointed instruments were unable to perform until the unit's replacement during the 1984 Space Shuttle Repair Mission.) The reaction wheel controlling the E~--~W motion would have spun up over a short time, moved uniformly for most of the excursion, then decelerated to a stop at the end of the E~--~W motion. The reaction wheel operating in the N~--~S direction would then have spun up over a short time interval, followed by a spinning up of the E~--~W reaction wheel in the reverse direction. Thus, another part of the raster pattern would have been formed, and so on till the raster pattern was complete, a total time of approximately 22 minutes. Because of the short but non-zero amounts of time for the reaction wheels to spin up, the motion would not have been exactly uniform. These non-uniformities were hardly apparent apart from a slight spectral smearing in some cases.

After the spacecraft scanning maneuvers, there was a period (22:37~--~22:57 UT, Phase~$C$) when the spacecraft attitude was undetermined and no meaningful BCS spectra were obtained. During a two-minute-long final phase before the sunlit portion of the orbit ended, starting at 22:57 (Phase~$D$), the spacecraft was pointed to another flare-productive active region, AR~2776, near disk centre, and remained there until the end of the orbit. A flare was also in progress in this region, and BCS spectra showed a displacement of only around 1~arcmin.

\section{BCS Spectra during the \SMM\ Spacecraft Scanning Maneuvers}
\label{sec:BCS_Sp_SMM_scans}

\subsection{Detection of BCS Crystal Non-uniformities}

The BCS spectra were collected in bins that are linearly related to wavelength. For BCS channel~1, bin numbers ran from 1 to 254; the range sensitive to solar X-rays was 33~--~220 while the end bins registered emission from the spectrometer's calibration source. The dispersion, calculated from the bin numbers corresponding to the peaks of prominent line features due to \caxix\ lines $w$ and $z + j$ (Table~\ref{tab:line_features}), is 0.3040~m\AA/bin. Using this and the mid-channel wavelength of 3.198~\AA\ \citep{rap77}, the observable wavelength range is 3.169~--~3.227~\AA\ for an on-axis source. Owing to a peculiarity of the data readout system, bin numbers in the raw data stream decreased with increasing wavelengths (see equation~(4) of \cite{rap17} giving the relation between the two). A more intuitive convention is used here in which BCS bin numbers $B_n$ increase with wavelength, and are related to those of \cite{rap17} ($B_{RSP}$) by $B_n = (253 - B_{RSP}) + 1$.

\begin{figure}
\centerline{\includegraphics[width=0.75\textwidth,clip=]{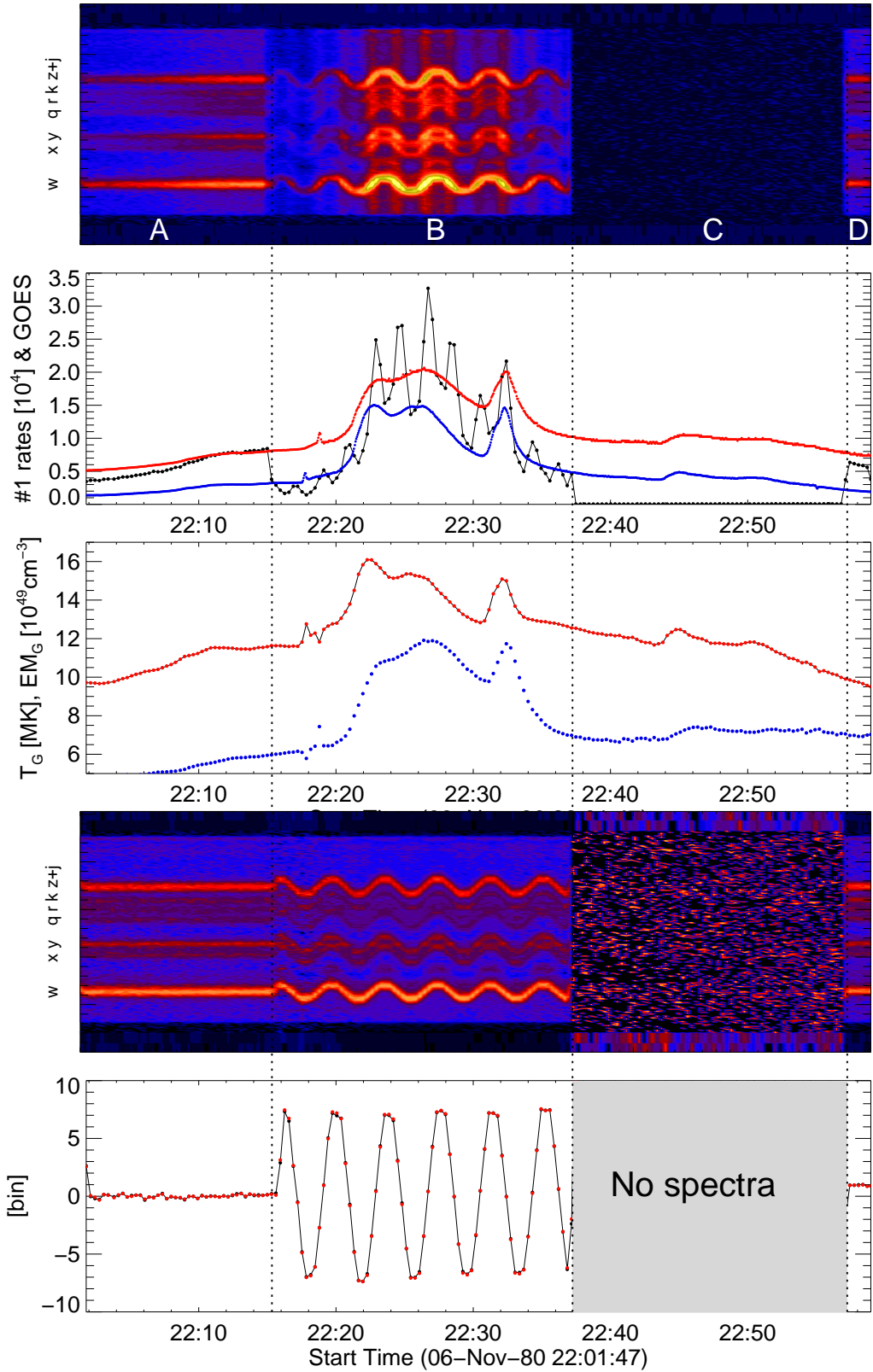}}
\caption{Time history of {\em GOES}\/, BCS Channel~1 (\caxix) spectra, and spectral line shifts during the 1980 November~6 flare (22:02~--~22:59 UT). Top panel: Channel~1 spectra plotted (vertical scale) on a red temperature intensity scale (yellow for high intensities, blue low). Second panel: BCS channel~1 photon count rates in bins 33~--~220 (3.169~--~3.227~\AA\ for an on-axis flare) in units of $10^4$~s$^{-1}$ with {\em GOES}\/ $0.5-4$~\AA\ (W~m$^{-2}$, multiplied by $2\times 10^5$, blue) and $1 - 8$~\AA\ (W~m$^{-2}$, multiplied by $6 \times 10^4$, red) light curves. Third panel: temperature ($T_{\rm G}$ in MK, red) and emission measure ($EM_{\rm G}$ in units of $10^{49}$~cm$^{-3}$, blue) from the emission ratio of the two {\em GOES}\/ channels. Fourth panel: Channel~1 spectra normalized to the BCS total count rate in the 3.169~--~3.227~\AA\ range. Bottom panel: wavelength shifts (expressed as relative bin number: 1~bin equals 0.3040~m\AA) determined by two different methods (red and black points: see text).
\label{fig:time_history}}
\end{figure}

Figure~\ref{fig:time_history} shows the time history of BCS channel~1 observations during the November~6 flare. BCS channel~1 spectra are shown as color intensity (red temperature scale) plots in the top and fourth panels with wavelength increasing from approximately 3.169~\AA\ to 3.227~\AA; the spectra in the fourth panel are normalized to the emission in this range. {\em GOES}\/ and BCS channel~1 light curves are shown in the second panel, and temperature ($T_{\rm G}$) and emission measures ($EM_{\rm G}$) derived from the intensity ratio of the two {\em GOES}\/ channels in the third panel. The four phases of spacecraft pointing ($A$, $B$, $C$, $D$) are indicated in the top panel. The {\em GOES}\/ light curves in Figure~\ref{fig:time_history} (second panel) show a number of peaks. Since {\em GOES}\/ has no spatial information, the sources cannot be unambiguously identified, but the BCS channel~1 total photon count rates have maxima that correspond to when the spacecraft was aligned with emission from AR~2779, so it can be presumed that the repeated maxima up till 22:35~UT are from AR~2779. The increase in the BCS count rate after 22:57~UT is due to the AR~2776 flare near disk center, which the spacecraft was pointed to just before spacecraft night.

The location of the BCS spectra are unaffected by spacecraft motion perpendicular to the dispersion axis, but for angular offsets in the plane of the dispersion axis, the spectra appear to ``slide'' back and forth along the detector as the Bragg angles of incidence vary on the curved crystal. The bottom panel of Figure~\ref{fig:time_history} illustrates the shift in bins of channel~1 spectra, in particular the oscillatory pattern made when the spacecraft performed its scanning motion during Phase~$B$. During the flare rise (Phase~$A$), some 41 spectra were obtained with the BCS optical axis pointed at the hot flaring kernel; prominent spectral line features, identified in Table~\ref{tab:line_features}, are indicated. During the spacecraft scans (Phase~$B$), 66 spectra were collected with the line features showing a wavy pattern reflecting the way the spacecraft E~--~W or W~--~E scanning motion moved the spectral lines along the dispersion axis -- apparent wavelengths increased as the spacecraft boresight moved toward solar east ($-x$) and decreased toward solar west ($+x$). In Phase~$C$, there are no recognizable features in channel~1 spectra. During Phase~$D$, with the spacecraft now pointed at AR~2776 near Sun center, channel~1 spectra showed no significant displacements, the flare being close to the BCS optical axis.

The shifts in BCS channel~1 spectra during Phase $A$ and $B$ (converted to bin units) were determined as follows: Each normalized spectrum in the bin range 33~--~220 was cross-compared with a normalized reference average spectrum to determine an optimum shift in BCS bins; this was done by interpolating the reference spectrum with a step size of 0.1~bin, moving each analyzed spectrum in steps of 0.1~bin to get the best overlap. Two criteria were applied: firstly, a ``multiplicative'' approach  (black points in the bottom panel) in which the shift corresponding to a maximum in the product of the analyzed and reference spectra was obtained; secondly (red points), the shift corresponding to where the traditionally defined $\chi^2$ difference attained a minimum value. The red points coincide with the black points to within 0.05~bin. Note that during Phase~$A$ the line features are not exactly stationary in bin space; there are slight fluctuations with a variation of $\sigma=0.14$~bin, representing the upper limit to the accuracy of the bin shift determination.

%

%
\begin{figure}
\centerline{\includegraphics[width=0.6\textwidth,clip=]{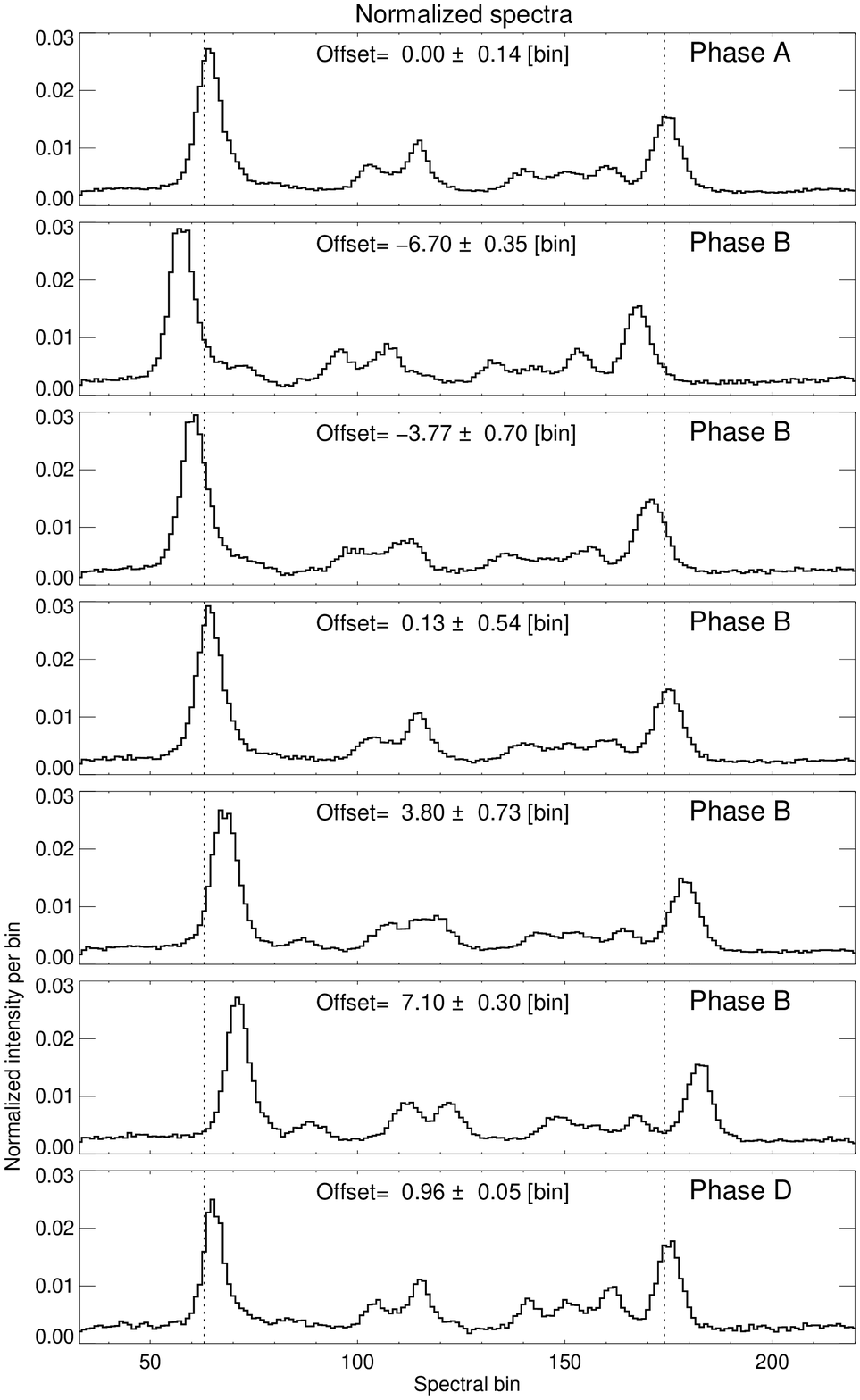}}
\caption{Normalized BCS spectra during the 1980 November 6 flare referred to BCS bin numbers. Top panel: average of 41 spectra in Phase~$A$ (zero offset). Second to sixth panels: Phase~$B$ spectra averaged over short intervals at averaged spacecraft offsets indicated in the legend (+ = offsets to spacecraft west; - = spacecraft east). Bottom panel: average of five spectra with \SMM\ pointed at AR~2776. The vertical dotted lines show the undisplaced bin positions of the \caxix\ lines $w$ and $z+j$.
\label{fig:Seven_BCS_sp_diffoffsets}}
\end{figure}

In Figure~\ref{fig:Seven_BCS_sp_diffoffsets}, spectral shifts in BCS channel~1 are illustrated by spectra plotted against bin numbers that are averages over seven periods. The first (top panel) corresponds to Phase~$A$, and is the average of 41 spectra during the AR~2779 flare rise (Phase~$A$), with a wavelength displacement of zero ($0.00 \pm 0.14$~bin). The following five panels show spectra during Phase~$B$, and illustrate how the spectra are displaced to shorter or longer wavelengths according as \SMM\ is pointed to respectively west or east of AR~2779 (offsets are indicated in each panel). The bottom panel shows averaged spectra during Phase~$D$ when \SMM\ was pointed at AR~2776 near disk centre. Spectra taken in intervals near the extreme ends of the spacecraft scans have slightly smaller line widths than those between the ends attributable to the fact that BCS were accumulated in data gathering intervals (19~s) that are significant fractions of the time that \SMM\ took to do a full E~--~W excursion or the reverse.

%
\begin{figure}
\centerline{\includegraphics[width=0.6\textwidth,clip=]{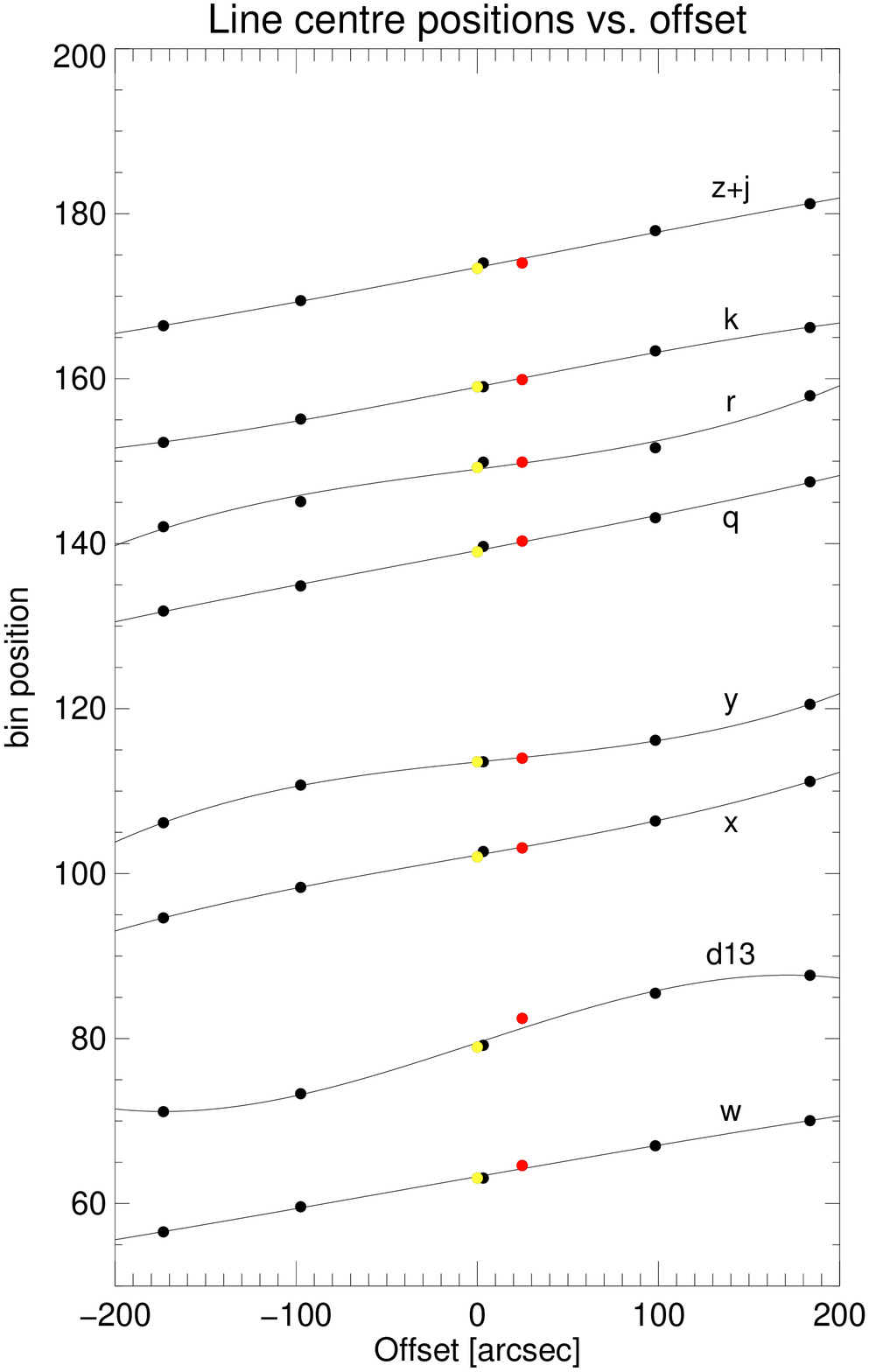}}
\caption{Measured bin positions of eight line features (labeled) in BCS channel~1 spectra, estimated by eye, plotted against spacecraft offset from boresight position in arcsec. The lines are: \caxix\ line $w$, \caxviii\ $d13+d15$ (labelled $d13$), \caxix\ $x$ and $y$; \caxviii\ $q$, $a$, $k$; and \caxix\ $z$ and \caxviii\ $j$ (blended). The black points are bin positions from spectra during the Phase~$B$ spacecraft scans; yellow points those from Phase~$A$; red points those from Phase~$D$.
\label{fig:linecenterpositions_vs_offset}}
\end{figure}

In Figure~\ref{fig:linecenterpositions_vs_offset}, displacement of eight recognizable BCS channel~1 spectral line features before (Phase~$A$), during (Phase~$B$), and after (Phase~$D$) the \SMM\ scan maneuvers are plotted against spacecraft offset (arcseconds). The line positions were determined by eye (automated routines were unreliable for weak line features like the $d13 + d15$ satellite feature).
There are no strong indications of any Doppler shifts in Phase~$A$ indicative of the flare impulsive stage, but even so we excluded Phase~$A$ observations from the determination of line feature positions (yellow points in Figure~\ref{fig:linecenterpositions_vs_offset}). Likewise, points in Phase~$D$ (red points) are excluded because they are from the AR~2776 flare. Thus, the points for Phase~$B$ (after the AR~2779 flare impulsive stage) are the most reliable. For five of the line features ($w$, $x$, $q$, $k$, $j+z$), the bin numbers of the points are distributed linearly with spacecraft offset. The bin numbers for two of the remaining five ($d13+d15$, $r$) line features are slightly non-linear with spacecraft offset, although these line features are relatively weak or partly blended. The points for the fairly strong line feature $y$ are also clearly non-linear. The non-linearity is most likely to be instrumental in origin, as discussed below.

\begin{deluxetable}{llr}
\tabletypesize{\small}
\tablecaption{Line features in BCS channel 1. \label{tab:line_features} }
\tablewidth{0pt}
\tablehead{
\colhead{Ion and line identification}  & \colhead{Transition} & \colhead{Wavelength(\AA)$^a$}\\}


\startdata
\caxix\ $w$                  & $1s^2\, ^1S_0 - 1s2p\,^1P_1$                                  & 3.17735 \\
\caxviii\ $n=3$              & $1s^2\,3p\,^2P - 1s\,2p\,3p\,(^1P)\,^2D$ ($d13 + d15$)        & 3.18180 \\
\caxix\ $x$                  & $1s^2\,^1S_{0} - 1s\,2p\,^3P_{2}$                             & 3.18941 \\
\caxix\ $y$                  & $1s^2\,^1S_{0} - 1s\,2p\,^3P_{1}$                             & 3.19291 \\
\caxviii\ $q$                & $1s^2\,2s\,^2S_{1/2} - 1s\,2s\,2p\,(^2P)\,^2P_{3/2}$          & 3.20048 \\
\caxviii\ $r$                & $1s^2\,2s\,^2S_{1/2} - 1s\,2s\,2p\,(^2P)\,^2P_{1/2}$          & 3.20332 \\
\caxviii\ $k$                & $1s^2\,2p\,^2P_{1/2} - 1s\,2p^2\,^2D_{3/2}$                   & 3.20663 \\
\caxviii\ $j$ + \caxix\ $z$  & \caxviii\ ($j$) $1s^2\,2p\,^2P_{3/2} - 1s\,2p^2\,^2D_{5/2}$    & 3.21095 \\
                             & \caxix\ ($z$) $1s^2\,^1S_{0} - 1s\,2s\,^3S_{1}$    \\
\\
\enddata
\tablenotetext{a}{From DIOGENESS: Phillips et al. (2018).}
\end{deluxetable}

\begin{deluxetable*}{llcll}
\tabletypesize{\small}
\tablecaption{BCS Channel 1 spectra and \SMM\ spacecraft scan motions. \label{tab:BCS_sp_times} }
\tablewidth{0pt}
\tablehead{
\colhead{Spectra no.}  & \colhead{Time range (UT)} & \colhead{Phase} & \colhead{BCS sp. offset } & \colhead{Spacecraft (S/c) Scan} \\
\colhead{} & \colhead{} & \colhead{} & \colhead{ch. 1 (bins)} & \colhead{E~--~W offset} \\
\colhead{} & \colhead{} & \colhead{} & \colhead{} & \colhead{$x$ in arcsec: W $= +$, E $= -$} \\
}
\startdata
1         & 22:01:38 - 22:01:57              & $A$   &                  & S/c repointing to AR~2779;\\
          &                                  &       &                  &  spectrum rejected.\\
2 - 42    & 22:01:57 - 22:15:08              & $A$   & $0.00 \pm 0.14$  & S/c not scanning; boresight \\
          &                                  &       &                  & at AR 2779 \\
43 - 45   & 22:15:08 - 22:16:05              &       &                  & S/c repointing to begin scans \\
46 - 111  & 22:16:05 - 22:37:00              & $B$   &                  & Spectra obtained every 19~s \\
          &                                  &       &                  & while S/c scans \\
  \\
46        & 22:16:05 - 22:16:24              & $B$   & +7.33            & S/c scans begin; S/c at min. $x$ \\
          &                                  &       &                  & ($x = - 174 $) \\
51        & 22:17:40 - 22:17:59              & $B$   & -6.97            & S/c at max. $x$ ($x = + 162$) \\
57        & 22:19:34 - 22:19:53              & $B$   & +7.18            & S/c at min. $x$ ($x = - 168$) \\
65        & 22:22:07 - 22:22:36              & $B$   & -6.71            & S/c at max. $x$ ($x = + 174$) \\
69        & 22:23:23 - 22:23:42              & $B$   & +7.02            & S/c at min. $x$ ($x = - 162$) \\
75        & 22:25:17 - 22:25:36              & $B$   & -7.02            & S/c at max. $x$ ($x = + 162$) \\
82        & 22:27:30 - 22:27:49              & $B$   & +7.40            & S/c at min. $x$ ($x = - 174$) \\
88        & 22:29:24 - 22:29:43              & $B$   & -6.74            & S/c at max. $x$ ($x = + 156$) \\
93        & 22:30:59 - 22:31:18              & $B$   & +7.20            & S/c at min. $x$ ($x = - 168$) \\
100       & 22:33:12 - 22:33:31              & $B$   & -6.60            & S/c at max. $x$ ($x = + 156$) \\
105       & 22:34:47 - 22:35:06              & $B$   & +7.50            & S/c at min. $x$ ($x = - 174$) \\
111       & 22:36:41 - 22:37:00              & $B$   & -6.32            & S/c at max. $x$ ($x = + 150$) \\
  \\
112 - 173 & 22:37:19 - 22:56:57              & $C$   &                  &  \\
174 - 175 & 22:56:38 - 22:57:16              &       &                  &  S/C repointing \\
176 - 180 & 22:57:16 - 22:58:51              & $D$   & $0.96 \pm 0.05$  & \\
181       & 22:58:51 - 22:59:04              &       &                  & S/C pointed at AR 2776 \\
\\
\enddata
\end{deluxetable*}

\subsection{Correction of BCS Spectra for Crystal Non-uniformities}

Long experience with the detector design, including laboratory development and pre-flight calibrations \citep{rap17}, assure that anode wire resistivity non-uniformities are not the origin of the channel~1 wavelength non-uniformities for \caxix\ line $y$, and instead they are to be ascribed to slight deformities in the crystal bend radii. Because of the latter, the bin widths do not exactly correspond to wavelength intervals, $\Delta \lambda$, and so the number of photon counts per bin does not quite correspond to photon counts per unit $\Delta \lambda$. The effect of deviations from the crystal's perfect curvature is to stretch or compress the spectrum with the same effective collecting area per detector bin, giving rise to intensity changes which therefore need adjusting.

These intensity changes affect satellite-to-line ratios which are important diagnostics of solar flare plasmas, as was discussed by \cite{gab72} and more recently summarized by \cite{por10}. For \fexxv\ and \caxix\ X-ray spectra, intensity ratios of dielectronic satellites such as $k$ and $j$ to the resonance line $w$ (Table~\ref{tab:line_features}: see also Figure~\ref{fig:three_BCS_sp}, discussed below) inversely depends on electron temperature $T_e$. With BCS channel~1 (\caxix) spectra, the relevant ratio is $k/w$ as satellite $j$ is blended with \caxix\ line $z$. For satellites $q$ and $r$, the upper levels of the transitions are primarily excited by electron collisions of the Li-like ion, and so the $q/w$ and $r/w$ intensity ratios allow the ratio of the Li-like ions to the He-like ions. This then enables checks on the presence or departure from ionization equilibrium in hot flare plasmas. For isothermal plasmas, significant differences in the derived temperatures from the ratios $(d13 + d15)/w$ and $k/w$ in \caxix\ spectra indicate the presence of non-thermal electrons in the excitation processes (see \cite{gab79}, relevant for \fexxv).

\begin{figure}
\centerline{\includegraphics[width=0.65\textwidth,clip=]{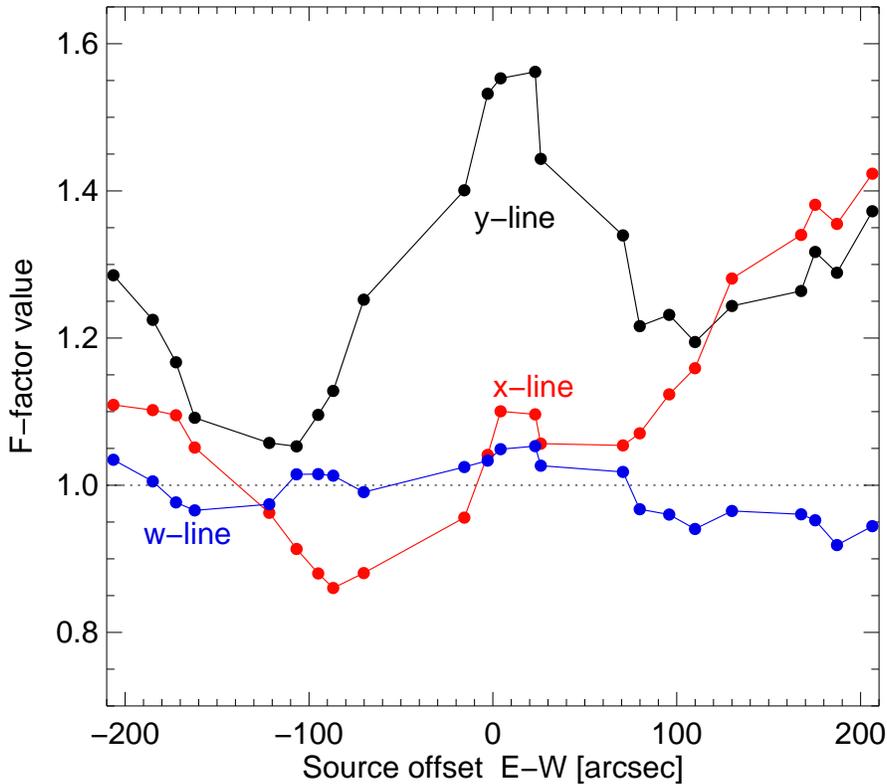}}
\caption{Values of correction factors $F$ for the intensities of \caxix\ lines $w$, $x$, and $y$ as a function of \SMM\ offset. Departures from 1 indicate anomalies in the crystal curvature. A large anomaly (1.54) is indicated for zero offset for \caxix\ line $y$.
\label{fig:corr_chan1_sp}}
\end{figure}

\begin{figure}
\centerline{
\includegraphics[width=0.32\textwidth,clip=,angle=0]{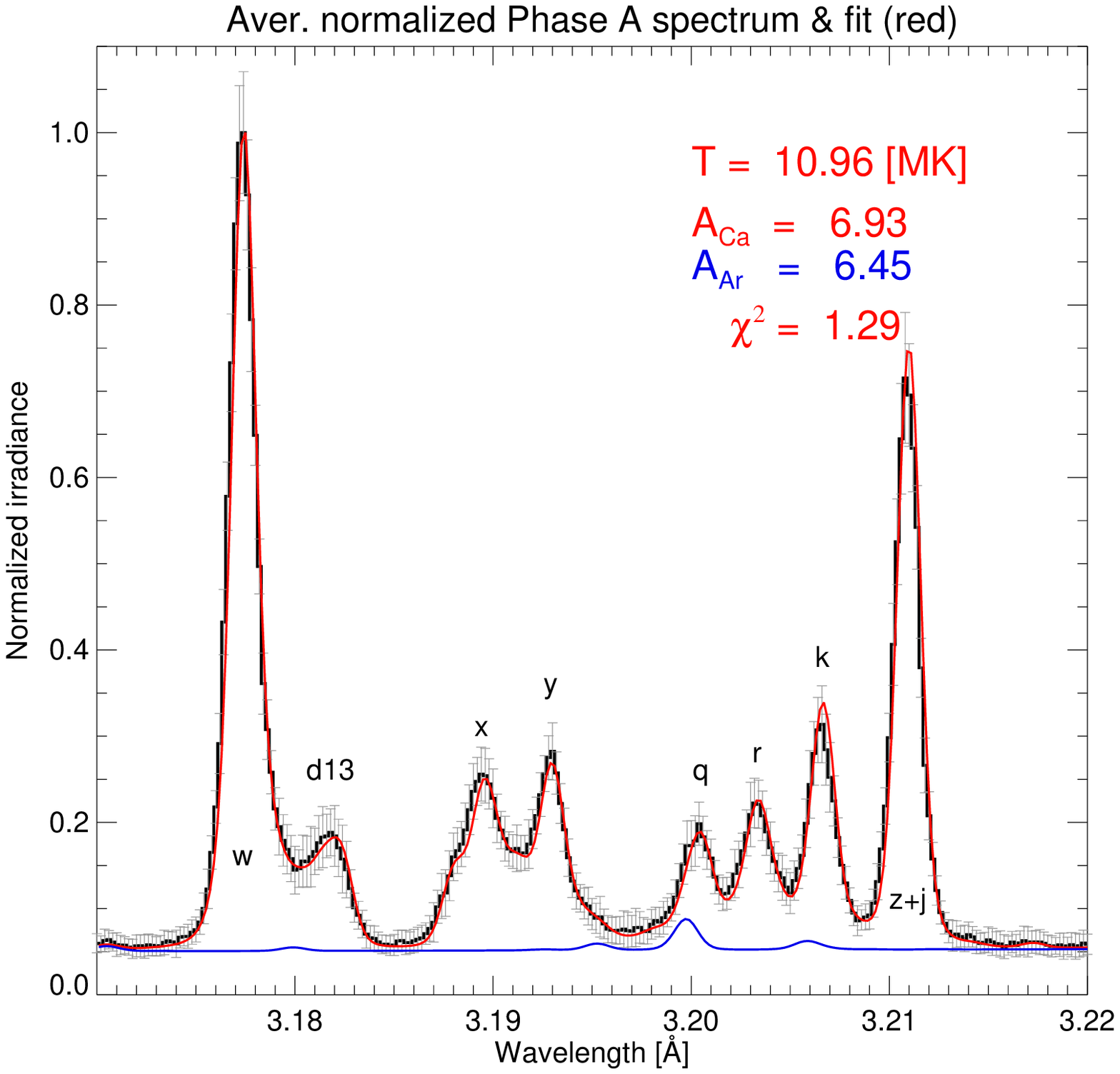}
\includegraphics[width=0.32\textwidth,clip=,angle=0]{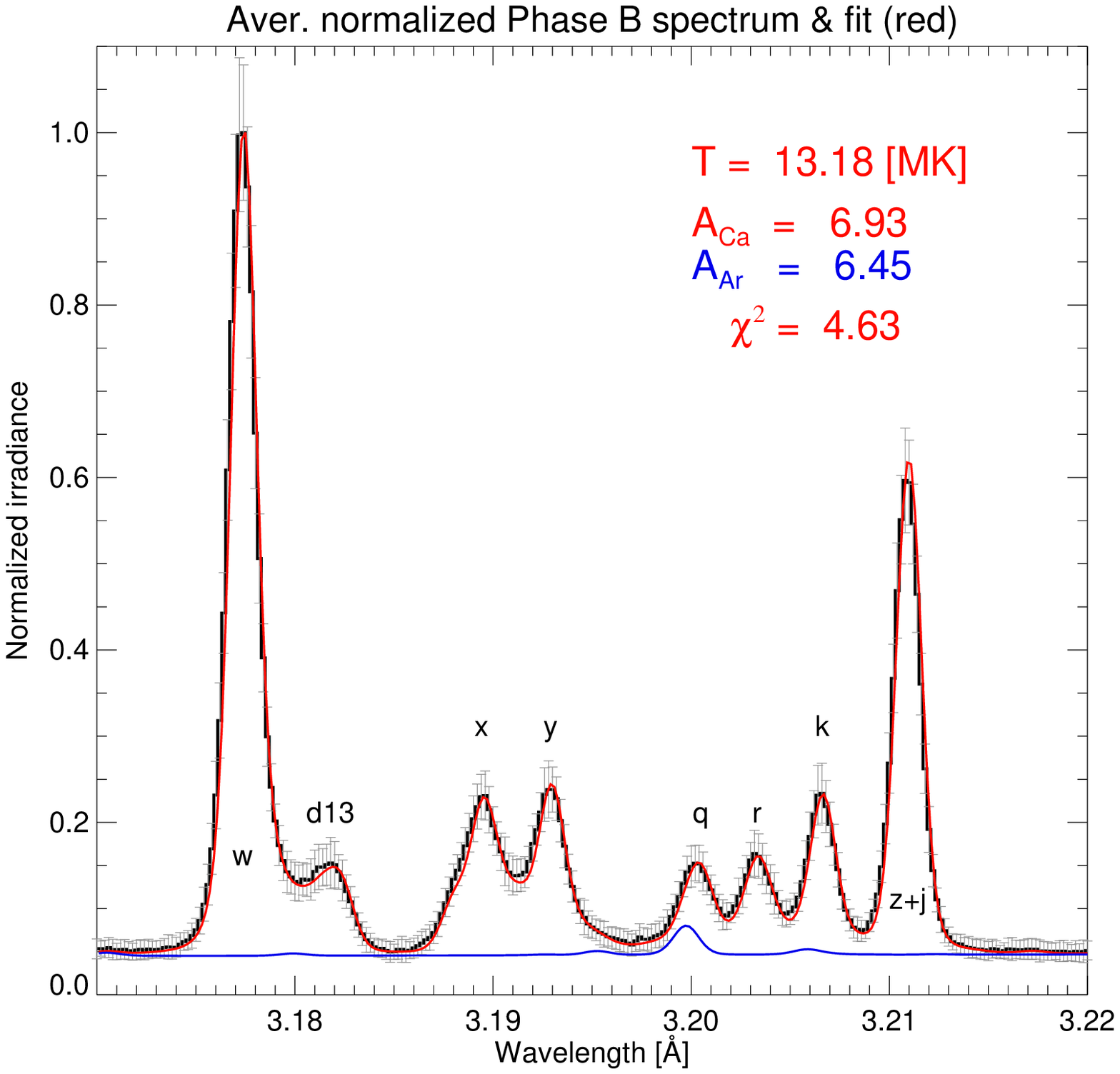}
\includegraphics[width=0.32\textwidth,clip=,angle=0]{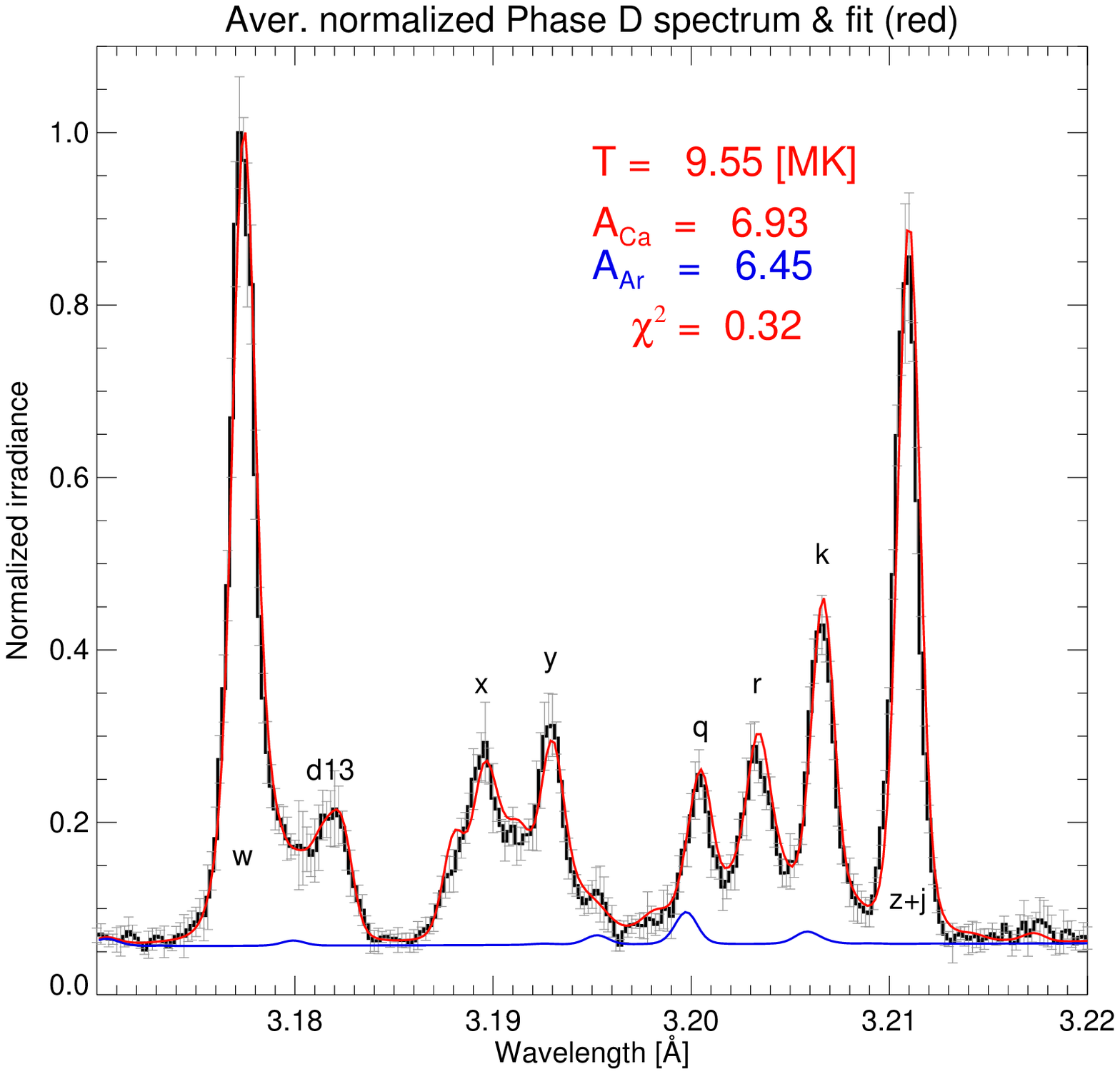}}
\caption{Normalized BCS channel~1 spectra (left to right) averaged over all spectra during Phase~$A$, Phase~$B$, and Phase~$D$ (shown as points with error bars representing photon count statistical uncertainties). The red curve is the theoretical spectrum (including continuum) as calculated by \cite{phi18}, and the blue curve the Ar~{\sc xvii} lines plus the continuum. The assumed Ca/Ar abundance is 3.0. Values of $\chi^2$ are indicated in each plot (the large value for the middle panel is due to the variation in temperature over Phase~B). \label{fig:three_BCS_sp}}
\end{figure}

To evaluate the intensity changes to the 112 BCS spectra collected during Phases $A$, $B$, and $D$ of the November~6 flare, we took each spectrum and found, using {\rm GOES}\/ data, the value of $T_{\rm G}$. As was established in an earlier work \citep{phi18}, this is nearly equal to the temperature from the dielectronic satellite-to-resonance line ratio $k/w$, or $T_{k/w}$ (see Table~\ref{tab:line_features} for identifications). We then defined factors $F$ for each wavelength giving the intensity ratio of each BCS spectrum to the calculated spectrum defined by $T_{\rm G}$, as described by \cite{phi18}. The matrix of $F$ values for Phase $B$ are a function of the offset of the \SMM\ boresight from the flare, and can be used to investigate crystal non-uniformities. It was found that there are non-uniformities at the wavelengths of the \caxix\ line $y$ and the relatively weak and partly blended satellites $r$ and $d13$. Non-uniformities at the position of line $y$ are of considerable importance as the intensity ratio \caxix\ $x/y$ has been problematical in previous analyses of BCS spectra, with line $y$ over-intense compared with theory (e.g., \cite{bel82b,phirainnie04}). This might ordinarily be attributed to additional atomic processes by which the upper level of line $y$ is populated. Even with improved collisional rates the discrepancy remained, and \cite{phirainnie04} concluded that it must be an instrumental effect, which is supported by the fact that the intensity anomaly has not been noted for other high-resolution solar flare spectra \citep{phi18} and laboratory spectra from the Alcator tokamak \citep{rice14}. The variation of the intensity ratio $F$ for the wavelengths of the \caxix\ lines $w$, $x$, and $y$ as a function of \SMM\ offset is given in Figure~\ref{fig:corr_chan1_sp}. While there is little variation for line $w$, there is a considerable variation for lines $x$ (large offsets) and $y$ (zero offset). The relatively large value (1.54) of $F$ for zero offset for line $y$ is of particular interest as it indicates an anomaly in the crystal radius at this location; BCS spectra at this wavelength should be divided by this value to obtain true intensity. Since the vast majority of flares seen by the BCS in its operational history were for zero offset, the intensity of the $y$ line needs to be corrected for all such spectra. 

%

The correction factors $F$ including those for \caxix\ line $y$ were applied to averaged spectra over Phase~$A$ (BCS optical axis directed at the AR~2779 M3.5 flare), Phase~$B$ (during the spacecraft scans), and Phase~D (flare in AR~2776). The corrected spectra are shown in Figure~\ref{fig:three_BCS_sp}. Error bars represent statistical uncertainties in the observed spectra. Each spectrum is compared with spectra (red lines) calculated as in \cite{phi18}. The calculated spectra are a function of $T_{\rm G}$ which as mentioned is nearly equal to $T_{k/w}$. The theoretical spectra in Figure~\ref{fig:three_BCS_sp} take into account the contribution of ionized Ar lines, in particular the Ar~{\sc xvii} $1s^2 - 1s4p$ line at the wavelength of the diagnostically important \caxviii\ $q$ line, as noted by \cite{dos81b}; the blue line in each spectrum shows the total contribution made by the Ar~{\sc xvii} lines plus the continuum. The abundance ratio of Ca to Ar is assumed to be 3.0, based on the average Ca abundance from \cite{jsyl98} and the Ar abundance from \cite{jsyl10b}. As can be seen from Figure~\ref{fig:three_BCS_sp}, the relative intensities of the \caxix\ $x$ and $y$ lines in the theoretical spectra are now very close to the observed, and therefore this well-known intensity anomaly is now explained by the non-uniformity of the BCS channel~1 dispersion.

The theoretical spectra include free--free and free--bound continua which were taken from routines in the {\sc chianti} database and code \citep{der97,delz15}. As can be seen from the three spectra in Figure~\ref{fig:three_BCS_sp}, not only the line emission but also the continuum in BCS channel~1 matches the theoretical continua seen in line-free regions to within only a few percent. This was assumed in analyses \citep{jsyl84,jsyl98} of the Ca lines in BCS channel~1 to obtain the flare abundance of Ca which was found to vary from flare to flare. The present work thus confirms the earlier work.

\section{BCS Collimator Angular Response}
\label{sec:BCS_Coll_Ang_Res}

In previous work \citep{rap17}, the BCS collimator transmission pattern was described as being of triangular shape in E~--~W ($x$) and N~--~S ($y$) directions with a nominal value of the FWHM equal to 360~arcsec and a peak transmission of 0.33 for a source on the BCS optical axis. The three-dimensional pattern was approximated by a pyramid shape in Figure~12 of \cite{rap17}, which is correct for the collimator transmission near the optical axis position but away from it ($\gtrsim 2$~arcmin) the pattern of the collimator transmission has a trapezoidal, not triangular, shape, and so requires a small correction. For distances in the E~--~W ($x$) and N~--~S ($y$) directions of the source from the optical axis, the transmission $C_T$ is given by

\begin{equation}
C_T = 0.33 \left( 1 - \frac{x}{{\rm FWHM}} \right) \left(1 - \frac{y}{{\rm FWHM}}\right).
\label{eq:BCS_trans}
\end{equation}

\begin{figure}
\centerline{\includegraphics[width=0.6\textwidth,clip=]{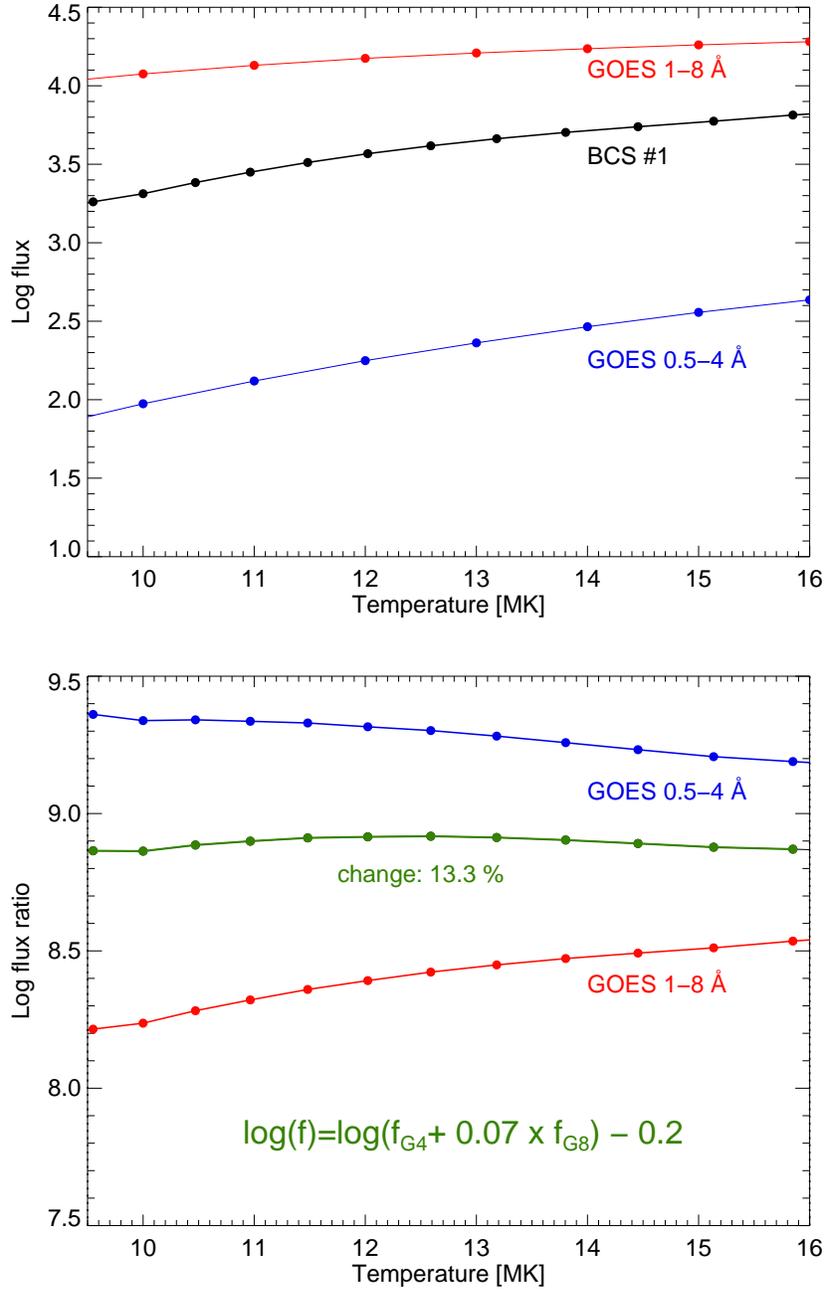}}
\caption{Upper panel: Temperature dependence of the logarithm of the flux from an isothermal source in the two {\em GOES}\/ channels ($0.5 - 4$~\AA, $1 - 8$~\AA, W~m$^{-2}$ multiplied by $10^8$ and $10^9$ respectively) and the total emission in BCS channel~1 shown as log (flux) with flux calculated for an emission measure $10^{48}$~cm$^{-3}$. Lower panel: The ratio of BCS channel~1 flux to $f_{G4}$ (blue points and curve), to $f_{G8}$ (red), and to the function $f$ where log~$f = {\rm log} (f_{{\rm G4}} + 0.07 \times f_{{\rm G8}}) - 0.2$ (green).
\label{fig:contrfuncts_GOES_BCS}}
\end{figure}

We used BCS channel~1 data during the 1980 November~6 flare to determine the in-flight transmission of the BCS collimator and obtain an empirical value for the collimator transmission FWHM. Rather than assume that $T_{\rm G} = T_{k/w}$ as earlier, we calculated the emission in the entire wavelength BCS channel~1 range (lines and continuum) and related it to the emission from an isothermal source in the two {\em GOES}\/ channels ($0.5 - 4$~\AA, $1 - 8$~\AA, called here $f_{{\rm G4}}$, $f_{{\rm G8}}$) separately. The emission is equivalent to the contribution or $G(T_e)$ curves that are generally given for individual spectral lines, and is defined here for a volume emission measure of $10^{48}$~cm$^{-3}$. The {\em GOES}\/ and BCS channel~1 curves are plotted in Figure~\ref{fig:contrfuncts_GOES_BCS} (upper panel). We empirically determined a function $f$, given by

\begin{equation}
{\rm log}\,\,f = {\rm log} (f_{{\rm G4}} + 0.07 \times f_{{\rm G8}}) - 0.2,
\label{eq:log_f}
\end{equation}

\noindent which over the observed temperature range of the November~6 flare ($9.5 - 16$~MK) is more nearly equal to the emission in BCS channel~1 than either $f_{{\rm G4}}$ or $f_{{\rm G8}}$. In the lower panel of Figure~\ref{fig:contrfuncts_GOES_BCS}, the ratio of the BCS channel~1 emission to $f_{{\rm G4}}$, $f_{{\rm G8}}$, and the $f$ functions are plotted against temperature; there is only a 13.3\% change in the ratio BCS channel~1 emission to $f$ over the $9.5 - 16$~MK range.

%

\begin{figure}
\centerline{\includegraphics[width=0.5\textwidth,clip=]{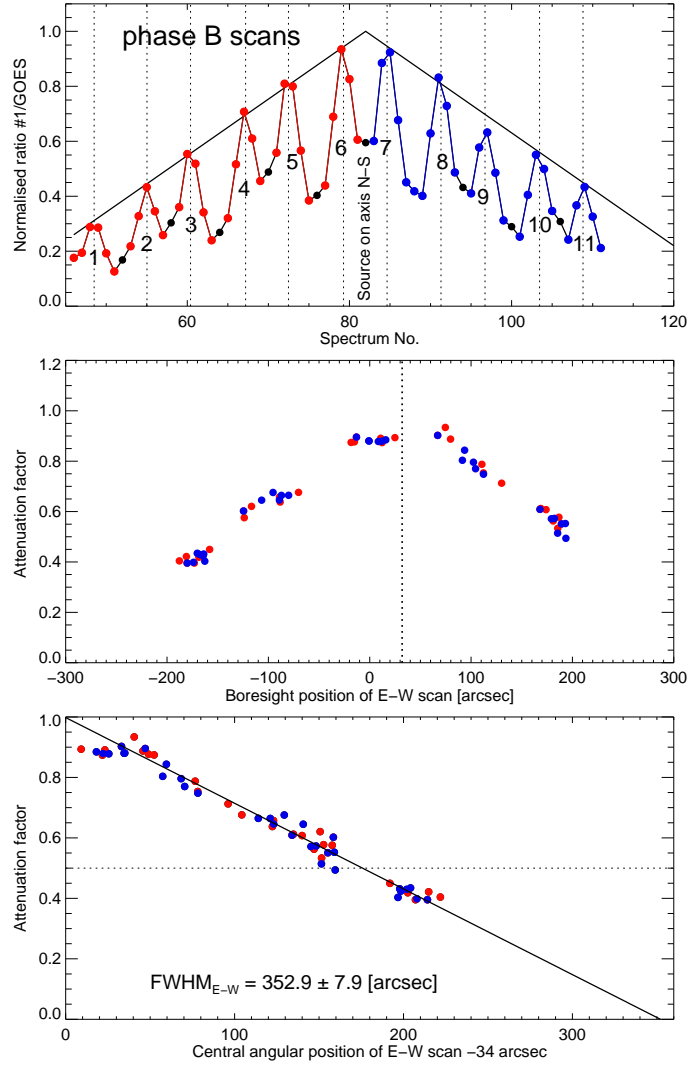}}
\caption{Top panel:  Ratio of emission in BCS channel 1 to the function $f$ plotted against the spectrum number (Table~\ref{tab:BCS_sp_times}). The ratios are normalized to one for a hypothetical source along the BCS optical axis. The diagonal lines are drawn through the maxima of corresponding E~--~W scans performed at fixed positions along the N~--~S direction. Center panel: Points from each of the E~--~W scans assembled using scaling factors corresponding to respective maxima (1~--~11) in the top panel.  Bottom panel: Points from left side of dotted line in center panel overplotted on the right side points. The straight line is the best-fit line defining the collimator FWHM in the E~--~W direction, determined to be $352.9 \pm 7.9$~arcsec. \label{fig:collimator_resp}}
\end{figure}

During the time period of the spacecraft scan (22:16:05~--~22:37:00), X-ray emission from the M3.5 flare varied considerably (Figure~\ref{fig:time_history}), so that emission at each time step of the scan is a convolution of the flare time variations and the movement of the spacecraft across the flare. However, having established that the BCS channel~1 emission is very similar to the {\em GOES}\/ emission function $f$, it is possible to reduce or nearly eliminate the flare time variations by plotting the ratio ($R$\/) of BCS channel~1 emission to $f$; a normalized version of $R$ is plotted in Figure~\ref{fig:collimator_resp} (top panel), and shows modulations as the BCS optical axis repeatedly passes either directly north or directly south of  the flare emission. There are eleven maxima (numbered in the figure) corresponding to times when the BCS optical axis was aligned with the flare. The progressive increase then decrease in the maxima is due to the BCS optical axis (and so spacecraft axis) scanning in an E~--~W or W~--~E direction. The direction of the spacecraft motion across the Sun in steps at the end of each E~--~W or W~--~E scan is determined to be towards the north from the fact that the BCS optical axis is to the south of the X-ray source (Figure~\ref{fig:sunspot_images}) before the scans began. In addition, the spacecraft scans (1 to 6) took a longer time to reach the X-ray source than to travel from it (scans 7 to 11).


Diagonal lines through the maxima in Figure~\ref{fig:collimator_resp} (top panel) meet at the time when the BCS optical axis was nearly exactly aligned with the M3.5 flare. For each of the eleven E~--~W scans, the ratio of the BCS channel~1 emission to $f$ (normalized to the ratio for a hypothetical source along the BCS optical axis) is plotted in Figure~\ref{fig:collimator_resp} (center panel). The points in this plot trace out the BCS collimator transmission along the dispersion direction. The central region is flattened which is probably due to the $\sim 30$~arcsec E~--~W extent of the flare source; this is indicated by the UVSP image in Figure~\ref{fig:sunspot_images} (green contours) which show that the flare extends $\sim 40$~arcsec in a N~--~S direction. In the bottom panel, both ``arms'' from the central panel are overplotted by mirroring in the dotted vertical line in the center panel. The width (FWHM) of the collimator response can be found from the slope of the diagonal lines, and has the value $352.9 \pm 7.9$~arcsec, only 2\% different from the pre-launch value of 360~arcsec \citep{rap17}.

%

\begin{figure}
\centerline{\includegraphics[width=0.5\textwidth,clip=]{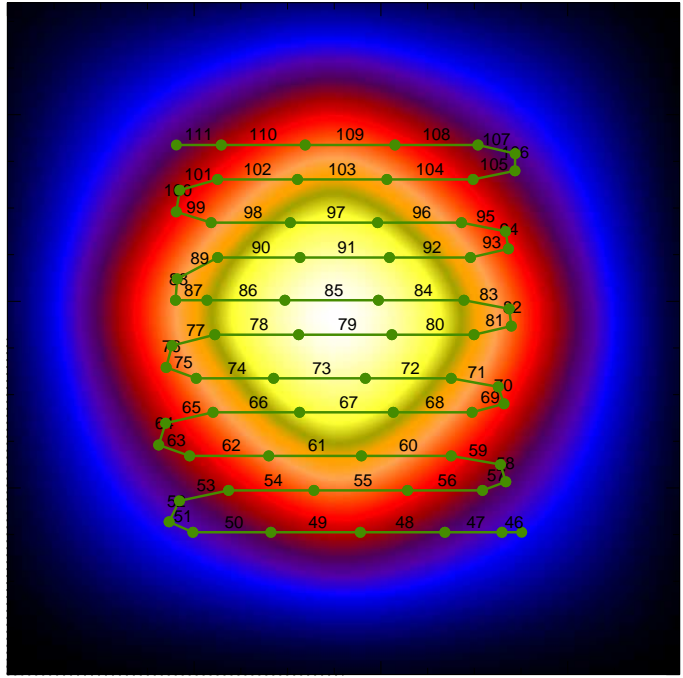}}
\caption{\SMM\ scanning pattern (green line) superimposed on the BCS collimator transmission response (colored background) with spectrum numbers shown (Table~\ref{tab:BCS_sp_times}).  \label{fig:SMM_scanning_pattern}}
\end{figure}

The \SMM\ scanning pattern deduced from the  BCS spectral shifts in the E~--~W direction, as well as the diagonal lines in Figure~\ref{fig:collimator_resp}, is shown in Figure~\ref{fig:SMM_scanning_pattern}. It is superimposed on the BCS collimator transmission response using Equation~\ref{eq:BCS_trans}. Spectra were taken over the intervals indicated by the numbers (see Table~\ref{tab:BCS_sp_times}).

In summary, the ratio of the BCS channel~1 total emission to the function $f$, which combines the emission in the two {\em GOES}\/ channels, allows the in-flight determination of the BCS collimator profile, and that the profile width (FWHM) is within 2\% of the pre-launch value.

%

\section{Summary and BCS Data Availability}
\label{sec:concl}

Archived data from the  \SMM\/ Bent Crystal Spectrometer are a unique resource for studying the X-ray spectra of the diagnostically important He-like Ca (\caxix) and Fe (\fexxv) line groups emitted by hot solar flare plasmas in that the spectral resolution was and remains (apart from the short-lived {\it Hinotori}\/ SOX2 spectrometer) the highest of any instrument, solar or non-solar. The BCS operated over the periods 1980 and 1984--1989, which included periods of very high solar activity (higher than any observed since) and through the solar minimum of 1986. In this work, using a data set during an M3.5 flare when the SMM spacecraft performed E~--~W scans over the active region, we measured the in-flight angular response of the BCS grid collimator. We used the fact that, as the BCS scanned over the flare, the flare intensity variations can be estimated and compensated for by taking the ratio of the total emission in BCS channel~1 to a function $f$ (Equation~\ref{eq:log_f}) that combines the emission in the two channels of {\em GOES}\/ rather than the BCS emission alone. The absolute intensity calibration, given by \cite{rap17} (Table 3), can now be refined for off-axis flares using Equation~\ref{eq:BCS_trans} with the collimator response width (FWHM) determined to be 352.9~arcsec.

Two other conclusions from BCS channel~1 \caxix\/ spectra have been drawn. The first is the finding that an apparent anomaly in the \caxix\ $x$ and $y$ line intensity ratios in BCS spectra is due to a non-uniformity in the Ge~220 crystal curvature, deduced from measuring the bin positions of prominent line features (Figure~\ref{fig:linecenterpositions_vs_offset}). The second is that the background emission in BCS channel~1 is almost entirely due to solar continuum observed in the portion of the Sun defined by the BCS collimator, so vindicating earlier work \citep{jsyl84,jsyl98} of flare-to-flare changes in the abundance of Ca. This indicates a more complex picture of observed variations in the abundances of elements with low first ionization potential than a simple multiple as has been previously suggested.

A few non-solar X-ray spectrometers have operated in the region covered by the \SMM\ BCS including the grating spectrometers on the {\em Chandra X-ray Observatory}\/ \citep{can05} and the microcalorimeter Soft X-ray Spectrometer (SXS) on {\em Hitomi}\/ \citep{hit18}. While these and other spectrometers have the advantage of a large spectral range, the spectral resolution is modest (resolving powers up to 1000) compared with the BCS for all its channels (resolving powers 4000 to 15000). The time resolution is, owing to the faintness of the targets, generally measured in tens of kiloseconds, so such instruments are unable to follow the temperature evolution of flares on active stars. A comparison of the HETG (High Energy Transmission Grating Spectrometer) spectra of two coronally active binary stars with solar flare spectra from the solar X-ray RESIK (Rentgenovsky Spekrometr s Izognutymi Kristalami) instrument on the {\em CORONAS-F}\/  spacecraft \citep{hue13a} shows the presence of He-like and H-like ion lines in the stellar spectra that are similar to those in solar flare spectra. The HETG spectra of both stars show the resonance lines of \fexxv\ and \caxix\ but only a hint of the satellite line structure. The deep exposures of the Perseus galaxy cluster reported by \cite{hit18} show several lines to the long-wavelength side of the \caxix\ and \fexxv\ resonance lines, some of which can probably be identified with dielectronic lines. Detailed analyses of these and similar line groups could possibly take advantage of solar flare BCS spectra as ``templates'' since the excitation of the lines is unlikely to be very different for particularly stars with X-ray-emitting coronae.

BCS data for the solar flares in the NASA \SMM\ archive\footnote{Description of the \SMM\ spacecraft and links to the BCS data archive are available from https://umbra.nascom.nasa.gov/smm/ } cover a range extending from small flares ({\em GOES}\/ rating B) to the very large flares seen in 1984 April, at the onset of Cycle~22. Software is already available for preliminary analysis of the spectra, written in the Interactive Data Language (IDL), with documentation. However, the corrections discussed here for the small deformations in the crystal curvature for BCS channel~1 and those needed for flares that were offset from the BCS optical axis have yet to be applied, but programming is underway. For BCS channels other than channel~1, viewing lines of highly ionized iron (wavelengths from 1.77~--~1.95~\AA), corrections for crystal fluorescence are needed which for some wavelength ranges are large and so estimates of the continuum may not be reliably determined. Work is in progress for these channels and will be the subject of future publications.

%
\acknowledgments

We acknowledge financial support from the Polish National Science Centre (grant number UMO-2017/25/B/ST9/01821). We thank Jaros{\l}aw B\k{a}ka{\l}a for his considerable help with the drawing of figures, and Craig Theobold (University College London Mullard Space Science Laboratory) for assistance in recovering original BCS drawings.

\facilities{Solar Maximum Mission (BCS)}
\software{SolarSoft Interactive Data Language \citep{fre98}, {\sc chianti} \citep{delz15}}

\newpage

%
\bibliography{RESIK}
\bibliographystyle{aasjournal}
%


\end{document}